\documentclass[preprint,12pt,authoryear]{elsarticle}

\usepackage{amssymb,url,amsmath}
\usepackage{enumitem}
\usepackage{float}
\usepackage{graphbox}
\usepackage[dvipsnames]{xcolor}
\DeclareMathOperator{\logit}{logit}
\journal{Computer Science and Language}

\begin{document}

\begin{frontmatter}

\title{A Study on the Manifestation of Trust in Speech}
 \author[icc,dc]{Lara Gauder\fnref{contributions}}
 \author[icc,dc]{Leonardo Pepino}
 \author[icc]{Pablo Riera}
 \author[unc,iip]{Silvina Brussino}
 \author[icc,dc]{Jazmín Vidal}
 \author[utdt]{Agustín Gravano}
 \author[icc]{Luciana Ferrer}

\address[icc]{Instituto de Investigación en Ciencias de la Computación (ICC), \\CONICET-UBA, Argentina}
\address[dc]{Departamento de Computación, Facultad de Ciencias Exactas y Naturales,\\ Universidad de Buenos Aires (UBA), Argentina}
\address[unc]{Facultad de Psicología, Universidad Nacional de Córdoba (UNC), Argentina}
\address[iip]{Instituto de Investigaciones Psicológicas, CONICET-UNC, Argentina}
\address[utdt]{Escuela de Negocios, Universidad Torcuato Di Tella, Argentina}

\fntext[contributions]{Lara and Leonardo contributed equally to the paper. Lara focused on the design, implementation and deployment of the data collection protocol as well as data curation. Pablo and Lara worked on the analysis of the collected database. Leonardo worked on the machine learning approaches, experimentation and analysis of results. Silvina helped design the data collection protocol. Jazmin helped with data curation. Agustín and Luciana directed the work, with hands-on contributions in the design of the protocol, the code, the machine learning approaches and the experiments.}

\begin{abstract}

Research has shown that trust is an essential aspect of human-computer interaction directly determining the degree to which the person is willing to use a system. An automatic prediction of the level of trust that a user has on a certain system could be used to attempt to correct potential distrust by having the system take relevant actions like, for example, apologizing or explaining its decisions. In this work, we explore the feasibility of automatically detecting the level of trust that a user has on a virtual assistant (VA) based on their speech. Since, to our knowledge, no public databases were available to study the effect of trust in speech, we developed a novel protocol for collecting speech data from subjects induced to have different degrees of trust in the skills of a VA. The protocol consists of an interactive session where the subject is asked to respond to a series of factual questions with the help of a virtual assistant. In order to induce subjects to either trust or distrust the VA's skills, they are first informed that the VA was previously rated by other users as being either good or bad; subsequently, the VA answers the subjects' questions consistently to its alleged abilities. All interactions are speech-based, with subjects and VAs communicating verbally, which allows the recording of speech produced under different trust conditions. Using this protocol, we collected a speech corpus in Argentine Spanish. We show clear evidence that the protocol effectively succeeded in influencing subjects into the desired mental state of either trusting or distrusting the agent's skills, and present results of a perceptual study of the degree of trust performed by expert listeners. Finally, we found that the subject's speech can be used to detect which type of VA they were using, which could be considered a proxy for the user's trust toward the VA's abilities, with an accuracy up to 76\%, compared to a random baseline of 50\%. These results are obtained using features that have been previously found useful for detecting speech directed to infants and non-native speakers. The collected speech dataset is publicly available for research use. 

\end{abstract}




\end{frontmatter}


\section{Introduction}
\label{sec:intro}

The ability to dynamically monitor the user's mental state, including their engagement, satisfaction, and emotions in general \citep{kiseleva2016predicting,sano2016prediction,kiseleva2017evaluating} is becoming an increasingly important component of conversational agents. 
In particular, and especially for virtual assistants (VA), tracking the user's degree of \emph{trust} in the system's skills may be critical for the success of the interaction \citep{parasuraman1997,drnec2016trust}. Distrust in a system can cause the user to under-use its capabilities. If a user starts displaying cues of distrust and the system can effectively detect such cues, then the dialogue manager could choose to act in consequence for regaining the user's trust \citep{muir1994,okamura2020,drnec2016trust}.

Several disciplines have investigated trust for decades, including psychology, anthropology, sociology, economics and political science. One important area of research has been the search for the factors that explain trust. Mayer et al.\ consider trust to depend on the trustor's perception of the trustee's \textit{ability, benevolence and integrity} \citep{mayer1995}. Other such factors have been proposed, including contextual and situational factors \citep{gulati1995,zucker1986}, and the propensity of a person to trust \citep{rotter1967}, among others. Trust has been also described as a dynamic phenomenon -- it can be created or destroyed during a conversation \citep{zand1972,korsgaard2014}. 

For systems that communicate with the user through speech, one possible way to track the level of trust is through the user's voice. It is reasonable to assume that a person would change the way they speak depending on whether they trust their interlocutor or not. Trust's nature has been depicted both as rational or cognitive \citep{coleman1990,hardin2006} and as emotional or affective \citep{miller1974}, or a combination of both \citep{mollering2006}. In either case, we hypothesize that the degree of trust affects and is affected by linguistic aspects (e.g.\ the form and content of discourse) and paralinguistic aspects (e.g.\ the intonation, pitch, speech rate and voice quality) of the trustor's and trustee's speech. The main goal of this research project is to study to what extent the trustor's degree of trust can be predicted from their speech signal during a human-computer interaction using fully automatic methods. We focus on eliciting and detecting trust or distrust of a person (the trustor) toward a VA's (the trustee's) abilities, leaving the study on the effect of benevolence and integrity for future work.

To date, very little research has been done in this area, but there are some indications that trust can be detected from the trustor's speech. \citet{waber2015} studied paralinguistic aspects in medical conversations between nurses and found that the emphasis used by an outgoing nurse when talking to an incoming nurse was significantly related to the degree of trust that the outgoing nurse reported to have on their colleagues. Further, \citet{elkins2013} found that the variations of pitch in time were related to the degree of trust of the speaker during human-computer interactions in the form of interviews. 

In this paper, we present a preliminary set of results that we hope will contribute to answer the question of whether trust can be detected from the trustor's voice. The experiments were conducted on the \textit{Trust-UBA Database}, which was collected specifically for this purpose, since, to our knowledge, no speech corpus was available with annotations of varying degrees of trust and large enough to allow for statistical analyses and machine learning experiments. Consequently, we designed and implemented a novel protocol for inducing varying degrees of trust in which subjects interacted with a virtual assistant (VA) in order to respond a series of factual questions. Before each series of questions, the subjects were told that the particular VA they were using had been rated with a high or a low score by previous users. This initial bias was later reinforced during the task by having the VA respond all or only some of the questions correctly, respectively. We subsequently used this protocol for building the \emph{Trust-UBA Database} in Argentine Spanish. We show results that indicate that the protocol succeeded in eliciting varying degrees of trust. Further, we show that a team of expert listeners achieved very low agreement in the annotation of the perceived degree of trust based only on the speech signal, indicating the difficulty of the task.

Using the Trust-UBA database, we investigated whether it is possible to automatically detect which type of VA the user is interacting with, a reliable or an unreliable one. 
To this end, we implemented a classification system based on a set of features extracted automatically from the user's speech. The features were motivated by the work done on speech directed to \emph{at-risk} listeners like infants, non-native speakers and people with hearing impairment, which we believed would share similar characteristics to speech directed to an unreliable VA. Research shows~\citep{hazan2015,scarborough2007,uther2007,saint2013motherese} that non-native- and infant-directed speech include some of the following characteristics: vowel hyper-articulation, a decrease in speech rate, an increase in the number and length of pauses, and an increase in pitch excursions. Further, work on speech directed to computers that make mistakes have been found to have similar characteristics \citep{oviatt1998}. Based on these works, we designed a set of features aimed at capturing these effects.

Our experiments on the Trust-UBA database using these features show that it is possible to detect whether a user is talking to a reliable or an unreliable VA based on their speech with an accuracy of up to 76\%. Note that we are not directly detecting mistrust but rather, a proxy given by the reliability of the VA. 

Yet, as discussed below, evidence indicates that users did trust the unreliable VA less than the reliable VA.
These findings suggest that we may indeed be able to detect the trust level from a user's speech, though further experiments are needed to confirm these findings on larger datasets with less controlled scenarios.

The rest of the paper is organized as follows. Section \ref{sec:protocol} describes and provides an analysis of the Trust-UBA database, Section \ref{sec:experiments} describes the machine learning experiments and results, and Section \ref{sec:conclusions} provides the conclusions.

\section{The Trust-UBA Database}
\label{sec:protocol}
In this section we describe the data collection protocol and the resulting database, and include an analysis of different aspects of the data. The full database is available for research purposes. Please, contact the authors if you are interested in obtaining the dataset.

\subsection{Protocol}

The protocol consists of an interactive session in which the subject must respond to a series of questions, aided by a speech-enabled virtual assistant (VA). A text-only version of this protocol was first described and evaluated in \citep{gauder2019}. In this section we describe the speech-based version of the protocol in detail. 

\subsubsection{Session structure and initial bias}
\label{sec:session_structure}
\begin{figure}[!ht]
  \centering
  \fbox{\includegraphics[width=.95\linewidth]{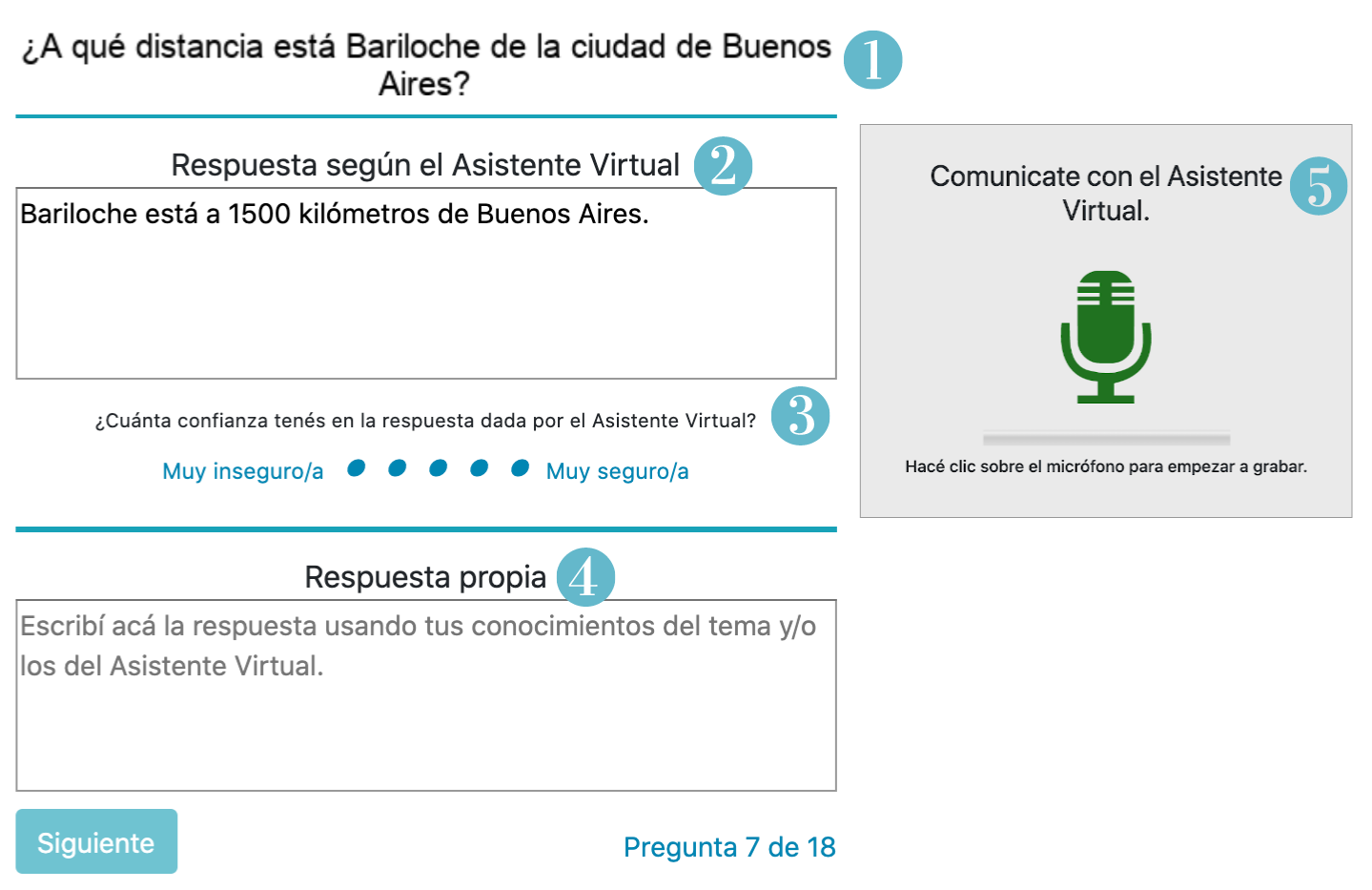}}
  \caption{Screenshot of the user interface for the question \textit{``What is the distance between Bariloche and Buenos Aires?''}\ (1). Subjects enter, from top to bottom: the answer according to the VA (2), their confidence in the VA's answer (3), and their own answer (4). Subjects interact with the VA using the voice recording button shown on the right: \textit{``Talk to the Virtual Assistant''} (5).}
  \label{fig:user-interface}
\end{figure}

The subjects' task is to respond to a series of factual questions with the help of a VA. For each question, subjects are required to
\textbf{(1)} interact verbally with the VA to find the answer to the question;
\textbf{(2)} listen and transcribe the answer given by the VA;
\textbf{(3)} rate their confidence in the response given by the VA using a 5-level Likert scale; and 
\textbf{(4)} enter their own answer, based on what they believe to be correct (this may or may not match the VA's response). 
Figure \ref{fig:user-interface} shows a screenshot of the user interface: the current factual question is shown at the top left of the screen; below that is a form in which subjects must enter the VA's response, their confidence in the VA's response, and their own response. On the right lies a voice recording button used to communicate with the VA.

At the beginning of a series of questions, subjects are told that the VA they will interact with was previously rated by other users with either a very high or very low score: 4.9 and 1.4 out of 5 stars, respectively (these two values were chosen empirically based on pilots tests; see \citeauthor{gauder2019}, \citeyear{gauder2019}). These two conditions are central in our protocol and are meant to bias the user toward either trusting or distrusting the VA's skills. We refer to them as the \textit{high-score} and the \textit{low-score} conditions, or simply H and L. 
With this setup we intend to benefit from a well-studied cognitive phenomenon called \textit{anchoring} or \textit{previous-opinion} bias, in which a person's decision-making process is influenced by an initial piece of information offered to them, such as a house valuation made by another broker, or a patient diagnoses made by another doctor \citep{sackett1979bias,tversky1974judgment}.
Subsequently, the quality of the responses given by the VA is consistent with the informed abilities, making no mistakes in the H condition, and making some mistakes in the L condition. This is intended to reinforce in the subject the feeling that the former system is good, and the latter is bad.

\subsubsection{Types of factual questions}

\textcolor{black}{For each condition, the subject solves a questionnaire that contains 18 questions which were chosen from a set of 36. In this paper, we will call each sequence of 18 questions corresponding to one condition from a recording session a \emph{series}.} When both conditions were completed by a subject, questions did not repeat across the two series. Hence, in those cases, subjects saw all 36 questions. Most questions appear in both conditions, though not exactly the same number of times, and 6 of the 36 question only appeared in the low-score condition. 

Of the 18 questions in each series, 6 were selected to be \textit{easy} and 12 to be \textit{difficult}.
Easy questions are about topics that should be obviously known by anyone (e.g., \textit{``How many days are there in a week?''}) and are used to generate the feeling in the subject that the VA actually works. Difficult questions, on the other hand, were selected so that their correct answers would likely be unknown to most people (e.g., \textit{``What are the three longest rivers in Argentina?''}). Thus, for difficult questions subjects should depend on the VA's responses. Furthermore, from the subjects' perspective, difficult questions make the task more challenging and interesting; and from our part, these questions allow us to manipulate the subjects' varying degree of trust in the VA's skills.

Difficult questions may be answered either correctly or incorrectly by the VA, as a reinforcement of the corresponding initial bias presented to the subject. In the L condition, 6 of the 12 difficult questions are answered incorrectly; in the H condition, all 12 difficult questions are answered correctly.
Importantly, no easy questions are ever answered incorrectly, since we found in pilot tests that doing so typically caused unnecessary frustration in the subjects, along with an irreversible feeling that the VA was useless. For that reason, incorrect answers to difficult questions were chosen to trigger a sense of \textit{doubt} in the subjects; even though they may not know the correct answer, they should feel that the VA's answer is wrong, without seriously hurting its reputation. For example, for the question, \textit{``What is the distance between Barcelona and Madrid?''}, the VA's incorrect answer is 1000 km (it is actually 504 km).

Lastly, questions can also be divided into two types, depending on the length of the interaction they are expected to trigger. Some questions and answers were prepared for forcing an exchange of several conversational turns, aiming at collecting more speech data from the subjects. For example, after the subject asks \textit{``What is the melting temperature of aluminum?''}, the VA may ask in which unit of measurement it should provide the answer (Celsius or Fahrenheit degrees). Using this strategy, we force subjects to have longer interactions with the VA and produce more dialogue acts -- not only questions but also answers. 

\subsubsection{Surveys}
\label{sec:evaluation-surveys}

At the beginning of each recording session, subjects are asked to complete a few questions regarding their demographic information, including gender, age, birthplace, and first and second languages. Next, they rate their self-perceived personality traits along 15 dimensions which can be mapped to the big five personality types, based on work by \cite{mondak2010personality}: 
creative, curious, intelligent, tidy, careful, hardworking, outgoing/enthusiastic/expressive, conversationalist, sociable, warm/affectionate, friendly, comprehensive, relaxed, calm, and stable.
Lastly, subjects rate their familiarity with, and trust in, virtual assistants and other digital systems, including GPS systems, automatic phone menus, and interactive-voice-response systems.

\begin{figure}[ht]
  \centering
  \includegraphics[width=.75\linewidth]{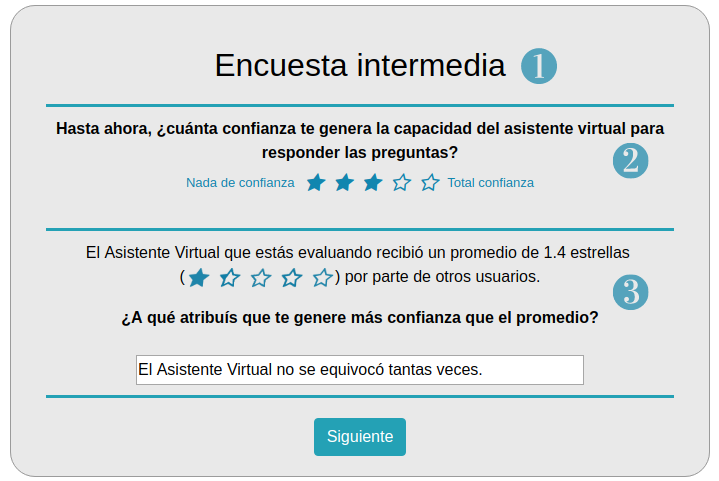}
  \caption{Screenshot of an \textit{intermediate survey} (1). Subjects answer two questions: \textit{``So far, how confident are you in the system's ability to answer questions?''}\ (2);  \textit{``The Virtual Assistant you are evaluating received an average of 1.4 stars from other users. Why do you think you gave a higher rating than the average user?''}\ (3).}
  \label{fig:va-survey}
\end{figure}

To assess the progress of the subjects' degree of trust throughout the series, they are required to complete simple \textit{evaluation surveys} after questions 6, 12 and 18. The first question, \textit{``So far, how confident are you in the system's ability to answer questions?,''} is answered in a 5-level Likert scale presented using a 5-star metaphor, as seen in the top part of Figure \ref{fig:va-survey}. Only after answering this question, subjects are reminded that the current VA received an average of $X$ stars by other users (as explained above, $X=4.9$ in the H condition, and $X=1.4$ in the L condition), and are required to explain in a few words why their score was higher or lower than the average (bottom part of Figure \ref{fig:va-survey}). 

The purpose of the evaluation surveys is twofold. They measure the progression of the subjects' degree of trust along the series, and they also refresh the anchoring bias introduced at the series beginning, thus reinforcing the H or L condition. The required textual explanation is intended to make subjects more conscious of this bias.

After completing a series of 18 questions and the third evaluation survey, subjects are required to rate how useful they found the VA, their degree of frustration with it, and how much they trusted it. They also report the extent to which they felt the following emotions and sentiments during the interaction:
active,
afflicted,
attentive,
tired,
decided,
disgusted,
distracted,
enthusiastic,
inspired,
uneasy,
nervous, and
fearful.
All questions are answered using a 5-point Likert scale.
These surveys are intended to further monitor and understand the subjects' behavior during their interaction with the VA.

\subsubsection{Implementation details}

Subjects interact with the VA via a voice recording button (right part of Figure \ref{fig:user-interface}), with which they may request the information needed to answer each question. The study interface was implemented online, to allow for data collection both in a controlled laboratory, and remotely over the Internet.

We built the VA dialogue system with the OpenDial toolkit \citep{lison2016opendial}, using a separate `dialogue domain' for each question -- i.e., a separate set of rules to trigger the system responses. We built a set of deterministic pattern-matching rules to cover the potentially many ways in which subjects may phrase their sentences.

We synthesized the VA's responses using Microsoft's publicly available speech synthesizer, with the HelenaRUS female voice in Spain Spanish with default settings.\footnote{https://azure.microsoft.com/en-us/services/cognitive-services/text-to-speech}
The subjects' utterances were transcribed using Google's publicly available automatic speech recognition system.\footnote{https://cloud.google.com/speech-to-text}

\subsection{Collected data}
\label{sec:data}
Subjects were recruited via ads on social media, emails to student mailing lists, and posters at the University campus. A total of 50 subjects participated at the University, in a controlled, silent environment (we call these the \textit{in-lab} subjects). This group was asked to solve two series of questions (one in each condition) and received a small monetary compensation for their time. Other 110 subjects participated over the Internet (these are the \textit{remote} subjects). In this case we had no control of the environment, which could result in poorer recording quality and lower concentration levels. This group was required to finish at least one series of questions, and were included in biweekly draws for a small monetary prize as compensation. 

For this paper, we decided to keep just the remote subjects that completed both study conditions, since much of our analyses and experiments relies on the comparison across conditions. Only 58 remote subjects completed both series of questions. From this subset, after listening to their audio samples, we discarded 24 subjects due to distorted signals, noisy environments, or too many network communication errors due to poor internet connections and other issues. \textcolor{black}{The remaining 34 speakers, along with the 50 in-lab speakers, are the ones released in the dataset and used for the analyses and experiments in this paper.}

From the 84 volunteers, 51 were female, 32 male, and 1 unreported. The mean age for the in-lab subjects was 24.26 (stdev 4.1) and 30.32 (stdev 12.08) for the remote subjects. All subjects reported Spanish as their first language. All but 2 in-lab subjects were born in Argentina, as well as all but 2 remote ones. Thus, the collected speech is overwhelmingly in Argentine Spanish.
The data corresponding to these speakers consists of 4930 short audios, with a mean duration of 4.05 seconds (stdev 2.1). The in-lab subjects recorded 2950 audios with a mean duration of 3.97 seconds (stdev 1.71). The remote subject recorded 1980 audios with a mean duration of 4.17 seconds (stdev 2.63). The average total duration of each session including both series was 49 minutes (stdev 12) and 56 minutes (stdev 39) for the in-lab and remote subjects, respectively. Most subjects did the second series right after the first one, with a single exception in the remote subjects, which did the second series a day after the first one. 

During some of the interactions, the speech recognizer or the dialog system implemented as part of the VA failed, causing the VA to ask the user to repeat the question. In previous works, it was shown that users change the way they speak when a system fails to understand their speech \citep{oviatt1998}, a phenomenon that could obscure the effect of the reliability of the VA. As we will see, we have taken this issue into account in our experiments. 

\subsection{Protocol effectiveness}
\label{sec:protocol_effectiveness}
\textcolor{black}{The main purpose of the protocol is to induce subjects into either trusting or distrusting the VA's skills. In this section we analyze its effectiveness, (1) by looking at the subjects' ratings given to the VA at each question and at the intermediate and final surveys, and (2) by comparing the answers subjects wrote based on the VA response and based on their own knowledge (boxes 2 and 4 in Figure \ref{fig:user-interface}).}

As explained in the previous section, occasionally, the VA made mistakes that caused it to ask the subject to repeat a question. All analyses in this section consider only the questions and surveys within a series that occurred before the first error made by the VA, since any responses from the subject after that are likely to be affected by the error. We assumed that the negative effect of a VA error in the subject's perception would carry over until a new series started and the user was told that this was now a new VA with a different score. Hence, questions in the second series (until the first error in the series) were included even if system errors occurred in the first series. In the case of the per-question analyses, we only included difficult questions that were answered correctly by the VA. We consider these questions to be the most relevant ones for our analyses, since we expect the rating given for easy questions and difficult questions answered incorrectly to be affected by the fact that the subject knows whether the answer to the question is correct or incorrect, and not by their level of trust toward the VA.
\textcolor{black}{Finally, for these analyses, we discard 39 subjects (19 in-lab and 20 remote of the 84 selected for the experiments in this paper) that experienced a system error before question number 6 and, in consequence, would have not answered any survey before that first error. Hence, the following analyses were made using the sessions from 45 subjects in total, 31 in-lab and 14 remote.}

\textcolor{black}{The code used for the analyses in this section is included with the publicly-available Trust database to ease replicability.}

\subsubsection{Analysis of question and survey ratings}
\label{sec:score_analysis}

\textcolor{black}{First, we used a Generalized Linear Mixed-Effects Model (GLMM) \citep{lme4, luke2017evaluating, lmerTest} to predict the condition of the series (H or L) given the score reported by subjects for each of the questions (star rating in Figure \ref{fig:user-interface}) including random effects given by the subjects. We assume that the logit of the probability that a certain question $i$ corresponds to condition H, given the score $s_i$ reported for that question and the subject identity $u$, is given by $\logit (P(\text{H} \,|\, s_i,u)) = a \cdot s_i+b_u+b_0$, where $\logit(p) = p/(1-p)$, $a$ is a global factor, $b_u$ is a subject-dependent bias, and $b_0$ is a global bias. To estimate the parameters of this model we use the \texttt{glmer} function from the \texttt{lme4} package in R.}
We found that the prediction can be made with high significance, $z\textrm{-value} = 4.07$ and $p = 4.61 \times 10^{-5}$, indicating that the scores reported by the subject were, as desired, affected by the condition of each series. 
A similar observation can be made when predicting the condition of the series given the scores provided by the subjects on the intermediate and final surveys (taken after questions 6, 12 and 18, see section \ref{sec:evaluation-surveys}). In this case, a GLMM with subjects and survey number (1, 2 or final) as random effects, gave $z\textrm{-value}=5.34$ and $p=9.05 \times 10^{-8}$.

\begin{figure}[!t]
\centering
\includegraphics[align=c,width=.49\linewidth]{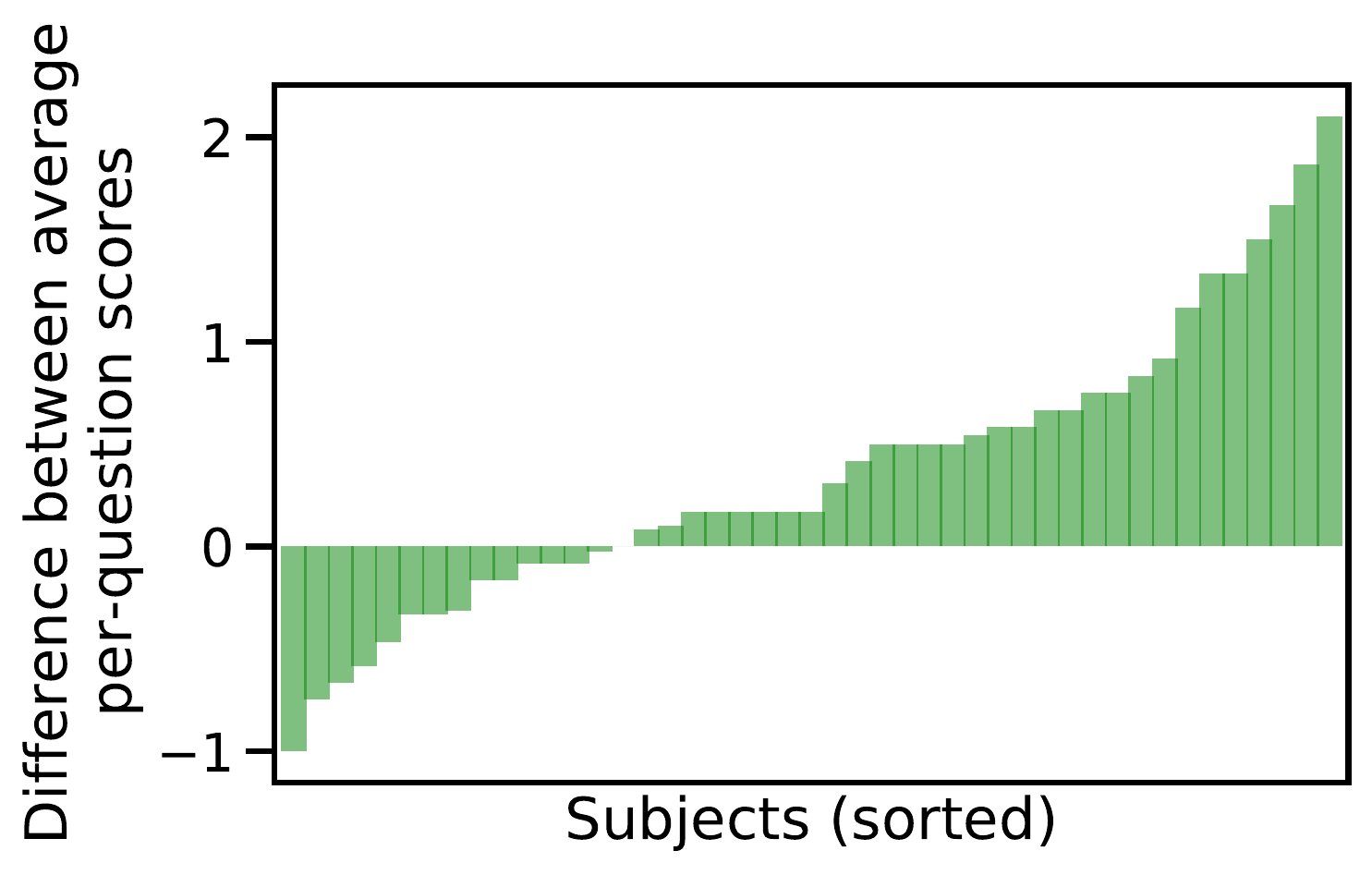}
\includegraphics[align=c,width=.49\linewidth]{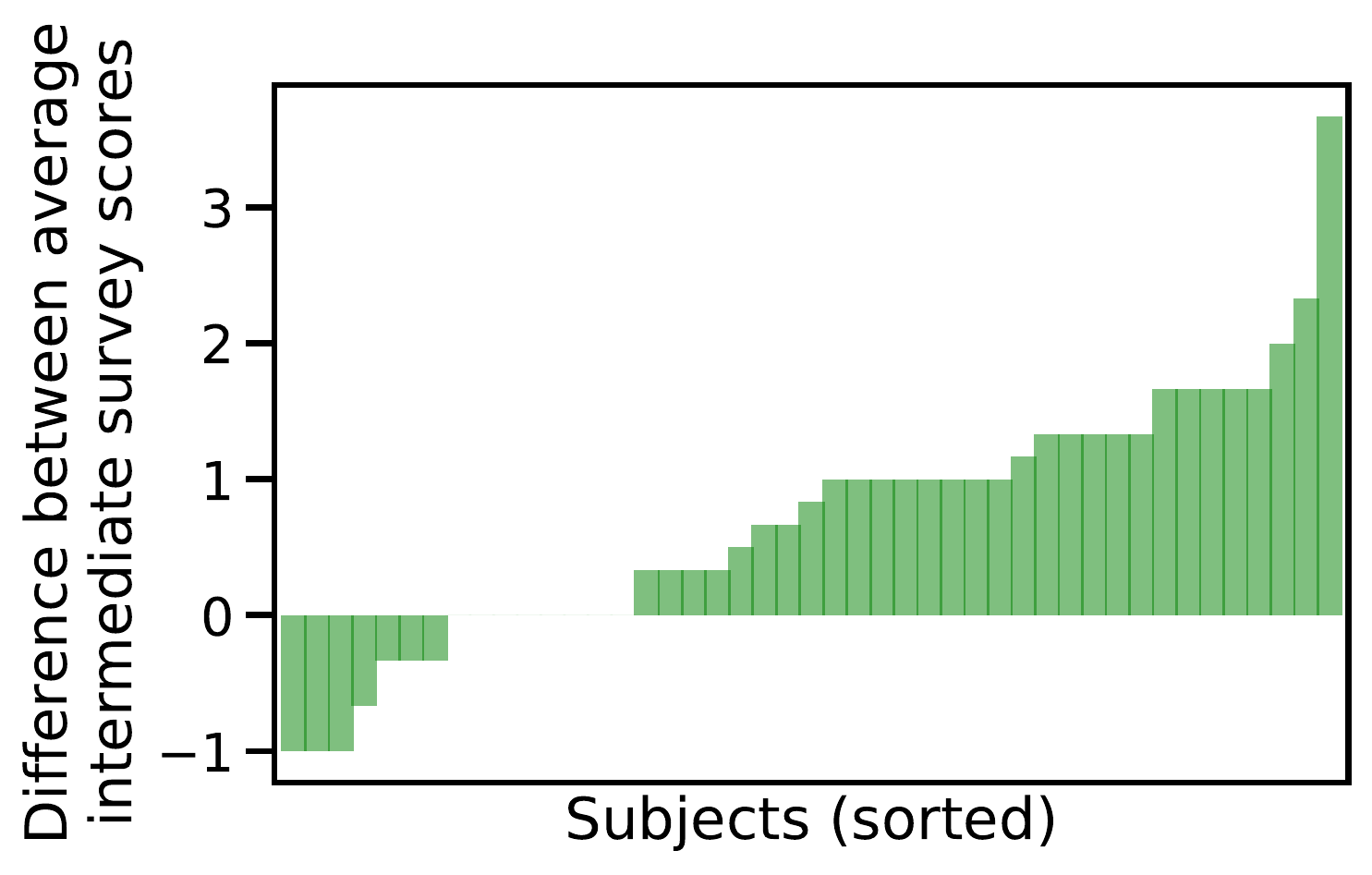}
\caption{Differences between average scores for the H and L condition for each of the subjects. On the left, per-question scores; on the right, intermediate survey scores.}
\label{fig:diff-trust}
\end{figure}

To visualize the effect captured by the GLMMs, we plot the difference between the average scores reported by the subject (before the first system error) for both conditions, on the per-question scores and on the intermediate survey scores. 
Figure \ref{fig:diff-trust} shows these differences for each of the 39 subjects with at least one intermediate survey before the first system error; the same samples used for the GLMM models. Subjects are sorted by the score difference. Positive values indicate the effect of the condition happened in the desired direction: a higher score was given, on average, to the more reliable VA. We can see that, for both types of scores, the majority of speakers was affected by the condition in the expected direction. 

\subsubsection{Analysis of the written answers}

\textcolor{black}{As we explained in Section \ref{sec:session_structure}, for each question, subjects were required to type the VA's answer and their own answer. By comparing both answers for a certain question we can get indirect information about whether or not the subject trusted the skills of the VA. That is, we expected that subjects would use the VA's answer for the difficult questions as their own answer more times for the H condition than for the L condition.}

\textcolor{black}{To analyze whether this expected behavior indeed occurred, we manually labelled the two answers (typed in boxes 2 and 4 in Figure \ref{fig:user-interface}) for a certain question as being either equivalent or different. We considered the answers to be the equivalent if the content but not necessarily the form was the same. For example, for the question \textit{``In which year was the first traffic light installed?''}, the answers \textit{``It was in 1868''} and \textit{``1868''} were considered equivalent, while \textit{``1868''} and \textit{``1890''} were considered different.
While doing this manual annotation, we found that 12 of the 45 subjects did not fill in their own answer with an actual response to the question, but rather used phrases such as \textit{``I don't know''} or \textit{``I agree''}. While for some of these answers one could infer whether the subject trusted or distrusted the VA, this was not always the case. For example, the answer \textit{``I don't know''} does not necessarily imply that the subject did not trust the VA's answer. Hence, we were forced to discard these speakers from this analysis.}

\begin{figure}[!t]
\centering
\includegraphics[align=c,width=.49\linewidth]{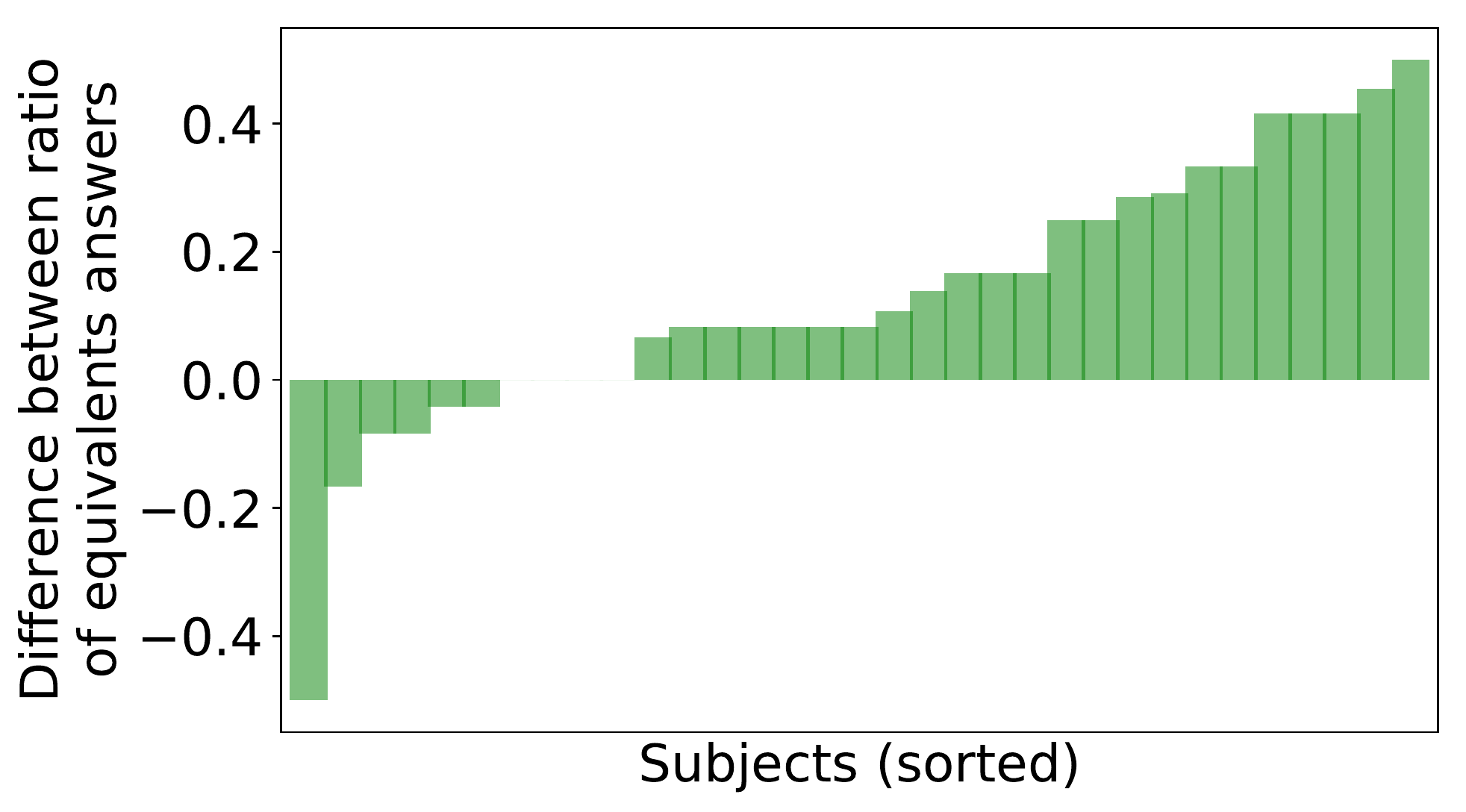}
\caption{Differences between ratio of equivalent answers given to  questions in the H and L conditions.}
\label{fig:diff-answers}
\end{figure}

\textcolor{black}{For the remaining 33 subjects we calculated the ratio of questions for which both answers were equivalent in each condition, considering (as explained above) only difficult questions answered correctly by the VA, and prior to the first error made by the system. Figure \ref{fig:diff-answers} shows the difference between these ratios for conditions H and L for each subject. We can see that, as expected, the subjects repeated the answer given by the VA as their own answer for a larger percentage of questions in the H condition than in the L condition.}

\textcolor{black}{In addition, we found a significant positive correlation between the difference between the means of the scores for each question for both conditions (corresponding to the y-axis in the left plot in Figure \ref{fig:diff-trust}) and the difference between the ratios of equivalent answers for both conditions (corresponding to the y-axis in Figure \ref{fig:diff-answers}). The Pearson correlation coefficient was $r(31)=0.498$ with $p\textrm{-value}=0.003$. This indicates that people that rated the VA as expected (higher for the H condition than for the L condition) also answered the question as expected (repeating the VA's answer more often for the H condition than for the L condition).}

\textcolor{black}{Overall, the analyses in this section and the previous one strongly support the conclusion that the protocol succeeded in eliciting the desired levels of trust.}

\subsection{Perceptual annotation of trust}

An additional research question in the current project is whether humans are capable of telling solely from the speech signal whether the speaker trusted or distrusted the VA's skills. We first conducted informal pilot studies in which the authors tried to perform this task, only to find it extremely difficult. We thus decided to gather a team of psychology researchers and practitioners who, as experts in human behavior, would be good candidates for succeeding in this task.

As a result, ten female expert annotators were asked to listen to a pair of sequences of audios from each in-lab subject (they thus listened to 50 pairs of audios). Each such sequence was formed by the first recordings produced by the subject for each of the final six questions in a series -- during which we expect the trust/distrust effect to be at its maximum level. The six audios in a sequence were merged into a wav file, separated by a simple tone. 
Each pair of sequences was presented to annotators on a web page, in random order. For each pair, they had to select which audio corresponded to utterances directed at the less trustworthy VA, together with their confidence level in a 5-level Likert scale. 
At the end of this study, annotators were asked to write in their own words what factors they considered when conducting this task. All annotators were paid for this task.

We examined inter-annotator agreement using Fleiss' $\kappa$ measure \citep{Fleiss71}, which yielded a value of 0.116. This is interpreted as ``slight'' agreement above chance. We also conducted a permutation test \citep{good2013permutation} (with 1000 permutations), which confirmed that this slight agreement is indeed significantly not random ($p \approx 0$). This suggest that the annotators did perceive certain speech cues related to trust, albeit faint and unreliable ones. An analysis of their written reports does not reveal any clear consensus. \textcolor{black}{As for the analysis in the previous section, the code used for this analysis is included with the Trust database.}

\section{Automatic Detection of Trust}
\label{sec:experiments}

In this section we describe our experimental setup and results for the task of automatically predicting the condition (high- or low-score, or simply H and L) from the subject's speech for each question or series of questions. Given the analysis presented in Section \ref{sec:score_analysis}, where we show that the condition of each series of questions significantly affected the level of trust reported by the subjects, the condition given by the initial bias for each series of questions can be considered a proxy for the trust that the subject had on the VA. Alternatively, we could have attempted to predict the scores reported by the subject (either on the intermediate surveys or per question). Yet, these scores are likely to be affected by hidden biases that are subject-dependent, since people may use the Likert scale very differently from each other. Hence, we leave the task of predicting the scores reported by the subjects for future work, and focus on predicting the condition. 

\subsection{Feature Description}
\label{sec:features}

As described in the introduction, the features used in this paper are motivated by previous work done on speech directed to at-risk listeners. In particular, we focused on the characteristics described by \cite{oviatt1998} as being related to hyper-articulation, which include more frequent and longer pauses, slower speech rate, clearer differentiation of vowel space with respect to formant values, increased pitch, and expansion of pitch range. The features described below aim to represent these characteristics using measures that can be automatically extracted from the waveforms. The only manual annotation done on this dataset is the orthographic transcription, which was first done automatically and then corrected manually when necessary. 

In order to detect the start and end of the speech for each utterance, and the duration of intermediate pauses, forced alignments to the manual transcriptions were performed using Montreal Forced Aligner \citep{montreal}. The extracted features were computed only over regions determined as speech by the forced aligner, and pauses shorter than 50 ms were considered part of the surrounding speech. Speech duration was calculated in two different ways: considering the full duration $T_F$ from the start to the end of the utterance; and also considering only the speech regions $T_S$, ignoring pauses between speech regions.

We computed a total of 16 features for each waveform, 3 related to speaking rate, 7 related to pitch, 2 related to energy, and 4 related to formant information.

\textbf{Syllable rates} including and excluding pauses, were calculated dividing the number of syllables by $T_F$ and $T_S$ respectively. The number of syllables in each utterance was calculated from the transcriptions using Syltippy,\footnote{https://github.com/nur-ag/syltippy} a Spanish syllabification tool based on \citep{hernandez2013}. \textbf{Pause to speech ratio} was also calculated as the total pause duration divided by the total speech duration $T_S$.

\textbf{Pitch features} were extracted using frame level estimates of the fundamental frequency (F0). F0 was calculated using OpenSmile's \emph{smileF0} configuration file, with an F0 tracking frequency range of 100 to 620Hz over frames of 50 ms shifted by 10 ms. OpenSmile assigns frames estimated to be unvoiced an undefined F0 value. The resulting F0 signals were further masked, turning all F0 values detected over pause regions into undefined values. The resulting signal was turned into a logarithmic scale and split into regions, defined as a sequence of consecutive frames separated by more than 50 ms of unvoiced frames. Finally, we computed the following summarized values: range, given by the difference between the 95th and 5th quantiles over all values; median over all values; mean and standard deviation over the regions of the median within each region; mean and standard deviation over the regions of the range within each region; and final slope, calculated using a linear regression over the last 25 (defined) frames of the F0 signal.

\textbf{Energy features} were given by the range and the slope over the last 25 frames for the energy signal extracted using OpenSmile. As for the pitch, this signal was restricted to have undefined values over unvoiced regions and turned into a logarithmic scale before computing the features.

\textbf{Formant features} were extracted for the first two formants. The formant estimates were obtained over voiced frames using OpenSmile and divided into regions as for the F0 signal.
For each of the formants we then calculated the mean and standard deviation of the ranges over the regions.

Note that we summarized the pitch contours using 7 functionals, but the energy and formant features using only a subset of 2 and 3 functionals, respectively. This selection arose from the characteristics described in \cite{oviatt1998}. We also tried summarizing the energy and formant features using the same 7 functionals applied to pitch, but this additional features did not improve classification performance. 

\subsection{Experimental Design}
\label{sec:design}
For the automatic prediction experiments in this section, we used only the sessions from Trust-DB recorded at the school laboratory, since they had better sound quality and fewer network communication and system errors. 
As for the analysis in Section \ref{sec:protocol_effectiveness}, we discarded all questions within a series that came after a VA's request to repeat a question due to ASR or dialog-system errors.  Further, for each question in the series, we used only the first waveform from the user that was not a mistake which required repetition due to the user's fault (e.g., stopping the recording before finishing the question). These waveforms had an average duration of 5 seconds.

To make the best use of the data, the experiments were done using a leave-one-speaker-out (LOSO) strategy, where a model is evaluated for each speaker using all the other speakers for training. The scores generated for all speakers using each corresponding model were pooled together to obtain one set of scores on the full dataset. 

The experiments were performed using a subset of 19 speakers for which at least 12 questions were available on each of the two conditions considering, for this count, only the questions (1) without transmission errors, (2) before the first system error in the series, and (3) that are not in the list of 6 questions that only appear in the L condition.
We considered two different tasks: 
\vspace{-0.1cm}
\begin{itemize}[leftmargin=*]
 \setlength\itemsep{0em}
    \item {\bf Question-level:} The unit of classification is each question within each series. In this case, the goal is to detect the condition (H or L) for the series in which the question is found.
    \item {\bf Series-level:} The units are the question-series. The ground truth in this case is the condition of the series. As explained later, for this case, the question-level features are summarized into series-level features for classification. 
\end{itemize}

\subsubsection{Feature Normalization}
\label{sec:norm}
The features described in Section \ref{sec:features} are likely to be highly affected by the speaker identity. This effect could potentially be more salient in the features than the condition we aim to detect. For this reason, we normalized the question-level features described in Section \ref{sec:features} over each subject's data, after filtering out questions with network communication errors or after the first system error in the series. \textcolor{black}{For normalization we use the standard score, also called $z$-score \citep{kreyszig}, where each feature is normalized by subtracting its mean and dividing by its standard deviation. Since the number of questions in each condition after filtering may be different, we used weighted statistics rather than standard ones, to avoid a per-subjects bias resulting from the imbalance in the conditions. This bias would hurt performance since, for some subjects the statistics would be computed with data mostly from condition H and, for others, with data mostly from condition L, making the resulting normalized features incomparable with each other.}

\textcolor{black}{We estimated the mean for feature $j$ for a certain subject $s$ as the weighted sample mean given by}
\begin{equation}
\label{eq:mean}
\textcolor{black}{\hat \mu_{j,s} = \sum_{i=1}^{N_s}{w_{s,i} \cdot x_{j,s,i}}\;,}
\end{equation}
\textcolor{black}{where $N_s$ is the total number of questions available for both conditions for subject $s$, $x_{j,s,i}$ is the value of feature $j$ for question $i$ from subject $s$, and the weight $w_{s,i}$ is given by $\frac{1}{2N_{s,c_{s,i}}}$, where $c_{s,i}$ is the condition (H or L) of question $i$ for  subject $s$ and $N_{s,c_{s,i}}$ is the number of questions for this condition available for subject $s$. Note that, with these definitions, $N_s = N_{s,H} + N_{s,L}$ and $\sum_{i=1}^{N_s} w_{s,i}=1$. The resulting mean is equivalent to computing the average of the by-condition means for the subject. The corresponding standard deviation was estimated using the same weights, as}
\begin{equation}
\label{eq:std}
\textcolor{black}{\hat \sigma_{j,s} = \sqrt{\sum_{i=1}^{N_s} w_{s,i} \cdot 
\big(x_{j,s,i}-\hat \mu_{j,s}\big)^2}\;.}
\end{equation}
\textcolor{black}{Finally, we normalized each feature by computing $z$-scores with  weighted statistics. Hence, the normalized features $\tilde x$ are given by} 
\begin{equation}
\textcolor{black}{\tilde x_{j,s,i} = \frac{x_{j,s,i} - \hat \mu_{j,s}}{\hat \sigma_{j,s}}. }
\end{equation}

Beside the user, another factor that could affect the features is the type of question. Some questions were simple to formulate (e.g., \textit{``Define the word bank''}), some were more complex or longer (e.g., \textit{``In what year did Hungary join the European Union?''}) and some were composed of two clear parts (e.g.,\textit{``Which are the three largest countries in the world, from high to low?''}) which most speakers divided up using a pause. For this reason, we explored the option of further normalizing the features by question. That is, after features are normalized by speaker, we computed the statistics per question over those features using only the training data. Those statistics were then used to normalize both the training and the test samples. This was done separately for each model being trained and the corresponding test samples for that model. 

The question statistics were also computed using weights to ensure that they were not biased by the imbalance between conditions for that question. 
For the six questions that appeared only in the low condition, weighted statistics could not be computed, since one of the conditions does not have samples. This was not a problem, since, as we will see, these questions were eliminated by the balancing process described below.

\textcolor{black}{As shown in Section \ref{sec:results}, this normalization procedure led to significant gains in performance with respect to using raw features.}

\subsubsection{Balancing Conditions for each Question when Training}
\label{sec:balancing}
Since each of the 36 questions used for the experiments did not appear the same number of times in the L and H condition, we undersampled each of the questions across the training data for each model to obtain exactly the same number of L and H cases. At this stage, the questions that appeared only in the L condition were discarded.
This balancing of condition per question avoids a possibly optimistic result where the model could be using the features to identify the questions, since the identity of the question would contain information about the condition. The undersampling is done randomly using 10 different seeds. For each seed, we ran LOSO and obtained a full set of scores on the test data. The scores obtained using different seeds were averaged obtaining a single score per sample in the test data.

Note that the balancing was only done during training, not during testing, since we only wanted to prevent the model from learning about the imbalance. During testing, all available questions (except the 6 that never appear in the H condition, since we do not have normalization statistics for them) were used to compute the summarized features, as described below.

\subsubsection{Classification Approach}

Classification was done using random forests \citep{hastie2001}, which are ensemble models designed to reduce the high variance of the estimations made by decision trees. Decision trees are very flexible models but, for this same reason, they are also very sensitive to small changes in the training set. Random forests, which are ensembles of such trees, reduce the variance and have been successfully used for many different classification problems \citep{hastie2001}. We also tried using support vector machines, which gave similar performance in terms of accuracy but which have the disadvantage of not providing a probabilistic measure at the output. Further, given the small amount of available data and the lack of held-out data, we did not want to try several different types of models or attempt to tune the different parameters in these models, since this could easily result in overly optimistic results.

We trained random forests consisting of 500 trees with a maximum depth of 20 using scikit-learn \citep{sklearn}. At each split, $\sqrt{N_F}$ features were randomly considered as candidates, where $N_F$ is the total number of input features. Splits were selected by minimizing the Gini impurity. These parameters were chosen based on our previous experience with random forests; they were not tuned to this dataset for the reasons mentioned above.

For the question-level experiments, the features input to the random forests were the features described in Section \ref{sec:features}, either raw, normalized by user, or normalized by user and question.
For the series-level experiments, we computed new features that summarize the distribution of each feature over the series. That is, given the questions within a series, we computed the 25, 50 and 75\% quantiles for each feature. This resulted in 48 summary features per series (16x3) which were used as input for the random forest classifiers.

\subsubsection{Performance Metrics and Bootstrapping}
We reported results in terms of cross-entropy and accuracy. The cross-entropy is given by $-1/N\sum_i \log(p_i)$ where $N$ is the number of test samples and $p_i$ is the posterior given by the system to the true class for sample $i$. We normalized the cross-entropy by the value it would have on a system that always outputs a 0.5 posterior, $\log(2)$. The accuracy is obtained over hard decisions made by thresholding the posterior returned by the random forests with a threshold of 0.5. 

We choose to report cross-entropy as well as accuracy because, while the accuracy is a standard and intuitive classification metric, it is quite noisy on small datasets like ours. A small change in the system output can significantly change the value of the metric. On the other hand, the cross-entropy is much more robust because it does not rely on hard decisions. For example, let's consider a comparison between two systems where the output changes for a single positive sample from 0.49 to 0.51. With that small change in output, this sample goes from being misclassified to being correctly classified. In our case, where we have 38 samples in total, the accuracy would change by 0.026, a relative improvement of 3\% if the accuracy of the first system is 0.76, as in our best system, which might tempt us to think that we have improved our system. Yet, this is probably not a change that would generalize to other data given the output changed only for one sample and only by a very small amount. Note that this would also be true for any metric that relies on hard decisions, like recall, precision or F1. On the other hand, for this same example, the cross-entropy would change by 0.00152, a relative improvement of 0.2\% for a initial cross-entropy of 0.83, as in our best system. This would tell us that the change between the two systems is, indeed, insignificant. 

Further, if the output for that same positive sample had changed from 0.499 to 0.8 (the new system is now very sure the sample is positive), the accuracy would still change by 3\%, but the cross-entropy would change 2.3\%, instead of 0.2\%, letting us distinguish these two cases. For this reason, we choose to show cross-entropy, alongside the more intuitive and standard accuracy values.

\begin{figure}[t]
    \centering
    \includegraphics[width=0.7\linewidth]{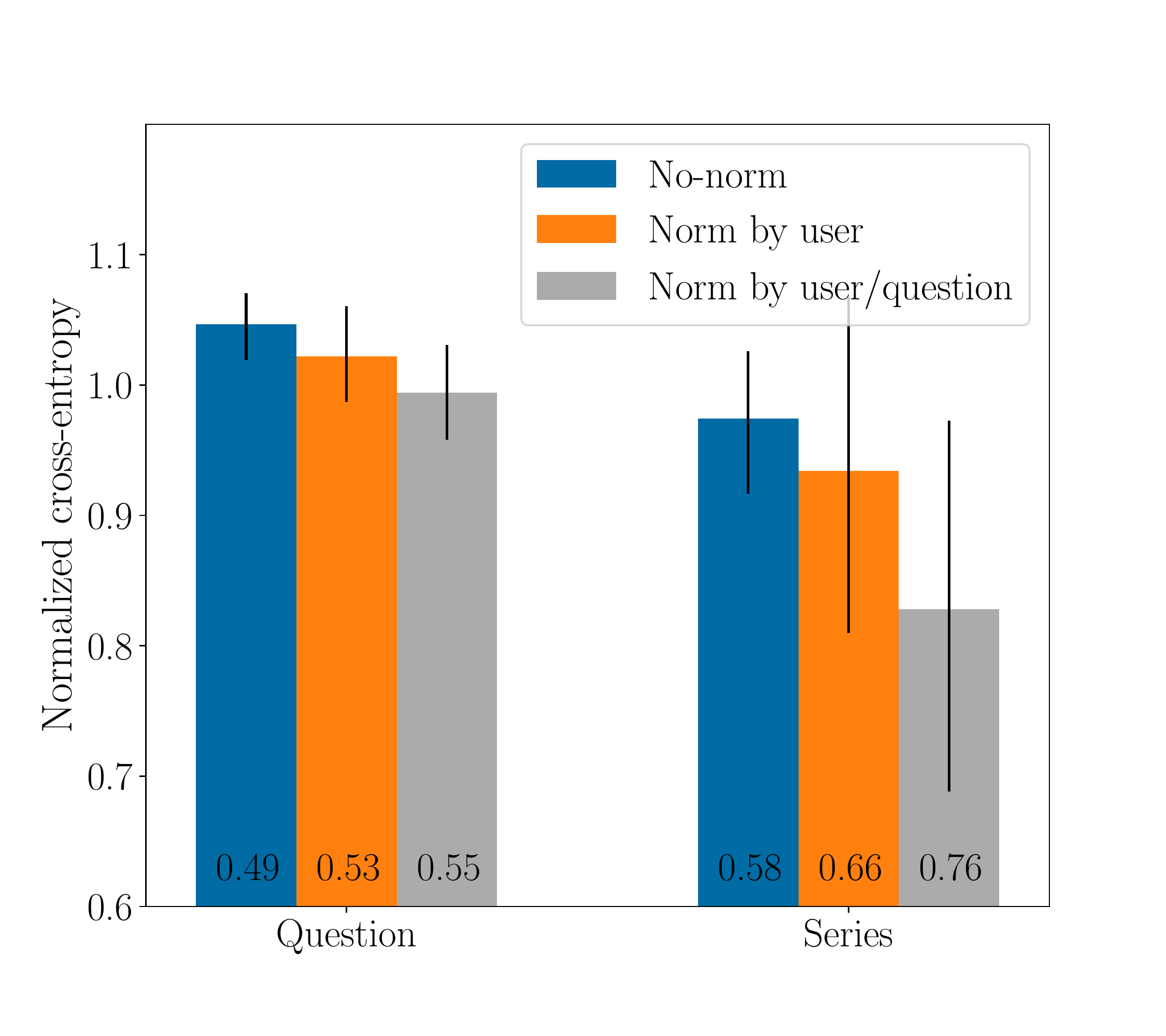}
    \caption{Normalized cross-entropy obtained at question and series levels for different normalization strategies. The height of the bars indicates the normalized cross-entropy, while the error bars are the confidence intervals obtained by bootstrapping. The numeric value at the bottom of each bar is the corresponding accuracy.}
    \label{fig:rf_results}
\end{figure}

\textcolor{black}{When evaluating a model, the value of the performance metric may depend on the test set size and composition.
In order to assess the uncertainty of our results, we obtained confidence intervals using  bootstrapping \citep{efron}. In this  method, multiple test sets are generated by sampling instances from the original test set with replacement. The performance metric is then calculated for each of the resulting test sets. These values can be used to derive a confidence interval measuring the variability of the metric of interest due to the test set. In this paper we use a version of the bootstrap method that considers the dependency across scores from the same speaker. As explained in \citep{poh:bootstrapping}, when the scores under evaluation come from a set of speakers with several instances each, the sampling should be done by speaker rather than by instance, so that the dependency across instances from the same speaker is taken into account. If this dependency was not accounted for, it could lead to an underestimation of the confidence interval. When sampling by speaker, for each bootstrap sample, some speakers might be missing and others might be repeated several times. The scores for the units (questions or series) for each speaker are discarded or repeated accordingly, and the resulting set of scores is used to compute a new estimate of the metric (cross-entropy, in our case).  The confidence intervals shown in Figure \ref{fig:rf_results} are given by the 2.5 and 97.5 percentiles of the distribution of cross-entropy values calculated with this method using 1000 estimates of the cross-entropy obtained by sampling the speakers with different random seeds.}

\subsection{Results and discussion}
\label{sec:results}

In this section, we report results for the model trained using features obtained at question and series levels applying the different normalization strategies discussed in Section \ref{sec:norm}.
Figure \ref{fig:rf_results} shows the cross-entropy and accuracy for the different tasks and normalization methods. We can see that, for both tasks, normalizing by user is better than not normalizing, and normalizing by user and question leads to the highest accuracy and lowest cross-entropy. These results indicate that, as suspected, features are affected by the user and the question, and removing these effects helps the model predict the type of VA more effectively.

For question-level classification, Figure \ref{fig:rf_results} shows that the performance is close to random, even after normalization is done (which, in effect, means that the features have information about the whole session). This indicates that just a few seconds of speech may not be enough to perform this complex task. On the other hand, the performance for series-level is better than random, indicating that the model is able to extract from the speech features useful information about the task of detecting the condition of the series. For the user- and question-normalized case, the accuracy of the series-level system was 76\% and the cross-entropy was 0.828, with a confidence interval that does not include the 1.0 value corresponding to a random system.

Speech features might be affected not just by the condition of the series but also by whether the person is getting tired or bored with the task. Hence, we analyzed whether the series-level predictions were affected by the order in which the two conditions occurred within the session (H followed by L, or the other way around). The scores given by the model to the 12 subjects who did the L series first had a mean of 0.574 (stdev 0.136), while the scores of the 7 subjects who did the H series first had a mean of 0.588 (stdev 0.104). The difference between these two distributions was not significant (Welch's $t$-test, $t(17)=-0.25$, $p \approx 0.8$). This suggests that the model is not using information related to the order of the series to make the prediction since, otherwise, we would expect the distribution of scores for both cases to be different. 

We applied recursive feature elimination to select a subset of the features (described in section \ref{sec:features}) that performed as well as or better than the full set. This analysis was done using the series as unit of classification, normalizing by user and question. First, we removed each of the features, one at a time, and calculated the cross-entropy in the validation set. Then, the feature that, when removed, gave the lowest cross-entropy was excluded from the training data and the procedure was repeated until removing features did not lead to lower cross-entropy values. 
Using this method we found that the best performing model needed a minimal set of three features: syllable rate including pauses, pitch median, and final pitch slope. With this minimal feature set, the random forest model achieves a normalized cross-entropy of 0.68 with a 95\% CI [0.47,0.92] and an accuracy of 0.76. The cross-entropy is improved by 38\% by this processing, suggesting that the correlation between many of the proposed features might be degrading the model performance.  

\begin{figure}[t]
    \centering
    \includegraphics[width=0.8\linewidth]{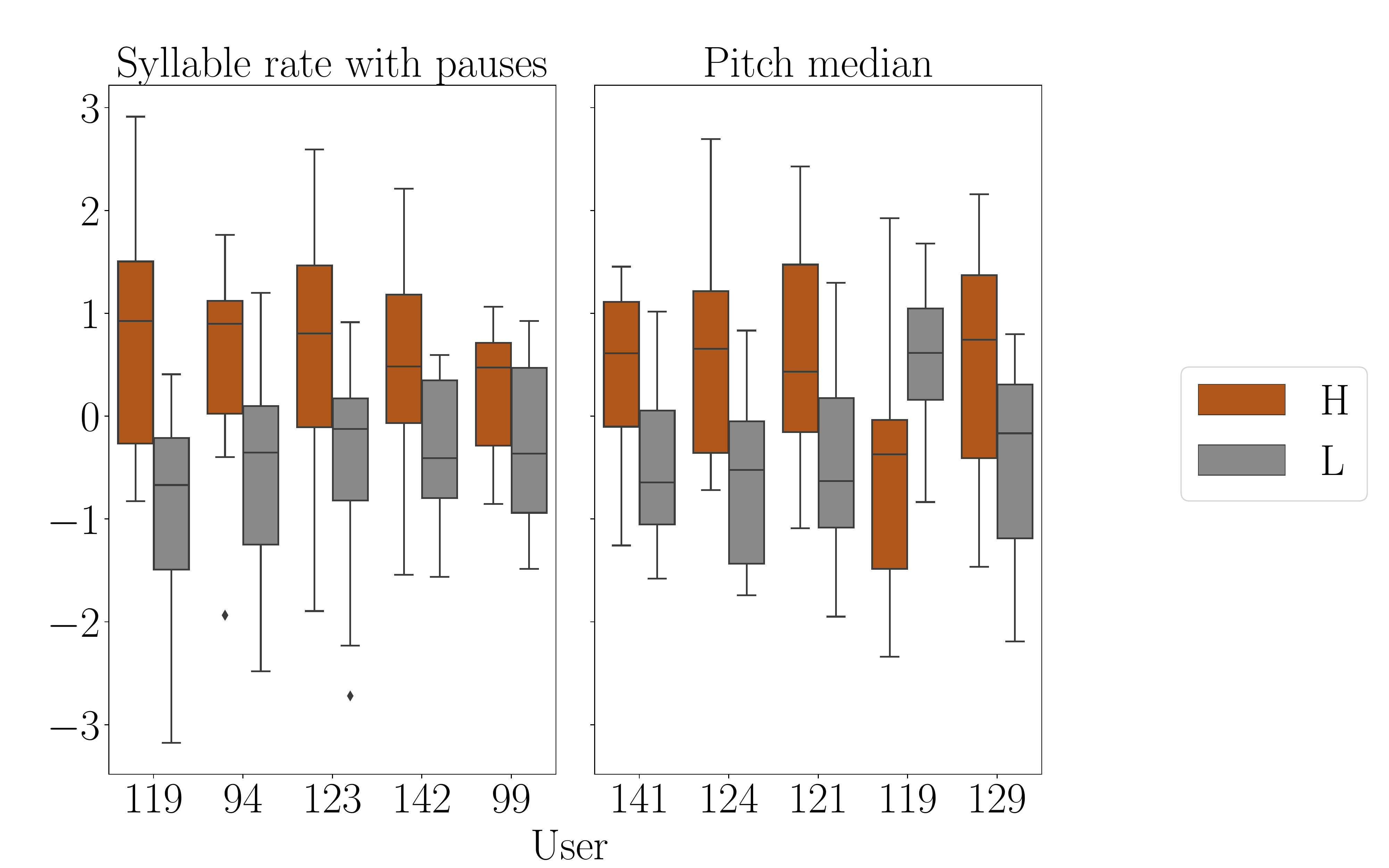}
    \caption{Box and whisker plots of the syllable rate with pauses and pitch median normalized by user and question for the 5 users with highest shift of the median between L and H conditions.}
    \label{fig:feats}
\end{figure}

Next, we analyzed the behavior of two top features selected by the recursive feature elimination method: syllable rate including pauses, and pitch median. We sorted subjects by their absolute difference in median across conditions for each of these features, and plotted the distributions for the five subjects that exhibit the highest differences (Figure \ref{fig:feats}). 
It can be seen that subjects that showed differences in syllable rate and pitch median tended to speak faster and with higher pitch in the H condition (with a single exception for the case of pitch). We also analyzed the distributions for final pitch slope, but did not find any clear patterns. We hypothesize that this feature may be used by the decision trees to condition the use of other features.

\subsection{Generalization to other subjects}

The experiments described above were performed on the 19 in-lab subjects that had at least 12 questions before the first system error (see Section \ref{sec:design}). We will call this set of subjects 12Q-in-lab.
As we saw in the previous section, the system performance on those subjects is significantly better than random.
Encouraged by these results, we decided to test the system on the in-lab speakers with at least 6 questions but fewer than 12 questions before the first system error and on the remote speakers with at least 6 questions before the first system error. We will call these sets 6Q-in-lab and 6Q-remote. They include 12 and 11 subjects, respectively. Results showed that the model trained on the 12Q-in-lab set did not provide good performance on these two other sets, achieving a cross-entropy of 1.033 for the 6Q-in-lab set and of 0.99 for the 6Q-remote set. These cross-entropy values around 1.0 indicate that the systems are not better than a random system that outputs a posterior of 0.5 for all samples. Further, retraining the models using LOSO on those new sets also resulted in poor performance. In this section we provide some hypotheses that might explain these results. 

The poor results in the 6Q-remote set were expected, since the remote sessions had a much less controlled environment than those recorded in our laboratory: subjects used different microphones, and had various background noises and distortion. Further, remote subjects were likely to be more distracted, less committed to the task. One symptom of this was the larger and more variable duration of remote sessions (see Section \ref{sec:data}) compared with in-lab sessions, indicating that subjects took breaks (in one case, a break of a whole night in between series). Finally, as we mentioned in Section \ref{sec:data}, the age range in the remote subjects was much wider than for in-lab subjects. In summary, the remote sessions were much more heterogeneous, making the task harder than for the in-lab sessions. We hypothesize that, to obtain a robust model under these more challenging conditions, we would need to train it on a much larger number of subjects.

We also expected results on the remaining in-lab subjects to be poorer than on the 12Q-in-lab set selected for our experiments, given that the amount of questions available for each sample (i.e., for each series of questions) before the first system error was smaller. To determine whether this was enough to explain the poor performance, we tested the models for the 12Q-in-lab subjects on a downsampled version of each of the test samples, where we kept only the first 6 questions in each series. The cross-entropy for this experiment was 0.885, a degradation compared to the 0.828 value we get when using all available test questions (Section \ref{sec:results}), but still better than the random performance we get on the 6Q-in-lab set. This suggests that the smaller number of questions per series is probably not the only reason why our model does not generalize to this set of subjects.

Another notable difference between the two sets is that the 12Q-in-lab subjects reported a higher difference in the average survey scores for the H and L series in the session $(M = 0.842, SD = 0.771)$ than the 6Q-in-lab subjects $(M = 0.069, SD = 0.641)$. This is a significant difference, with $t(29)=3.017, p=0.005$. 
Interestingly, on the 12Q-in-lab subjects, we found that there is a moderate but significant correlation ($r(17) = 0.46, p = 0.049$) between the average posterior given by the system to the true class for a session (obtained as the average posterior for the true class for the two series in the session) and the difference in the average survey scores that the user gave the VA for the H and the L series in the session.
That is, the system appears to label correctly with more confidence the subjects that reported being more affected by the reliability of the VA. This suggests that those subjects who did not report a difference in scores between the two series also did not change the way they spoke to the VA, thus making the system ineffective for those cases. This could, in turn, help explain the poor performance of the system on the 6Q-in-lab subjects, given that they had a significantly smaller difference in survey scores across conditions than the 12Q-in-lab subjects.

Note that the difference in scores was not the criteria used to select the subjects used in our main experiments. Instead, as described before, we selected the subjects with at least 12 questions per series before the first system error. Yet, there seems to be a correlation between the number of errors made by the system and the difference in scores.  Another interesting observation is that women reported a significantly larger difference in scores between the two conditions $(M=0.735, SD=0.703)$ than men $(M=0.167, SD=0.820)$, $t(30) = 1.945, p=0.061$, and also had fewer system errors. We hypothesize that subjects that were more committed to the task, a majority of them being women, also spoke more clearly (though not necessarily slower) to the VA.

\section{Conclusions}
\label{sec:conclusions}

We have presented results on the prediction of a proxy for the trust that a user has on a virtual assistant (VA) during a dialog, based on the user's speech. The proxy is given by the reliability of the VA (high or low) which correlates with the user's trust in the VA. 
Our task is, given two series of speech waveforms from the user recorded under both conditions (high or low reliability), predict the condition for each series. We show that a system can learn to perform this task with an accuracy of up to 76\%, where a random baseline would have a 50\% accuracy. To solve the task we designed features that are related to those previously found to be useful for detecting speech directed to ``at risk'' listeners such as infants, non-native speakers, or people with hearing loss.

We would like to emphasize that the experimental design used in this paper does not correspond to a realistic use case, since it assumes that data from both conditions are available for each user during training and testing. This setup was selected for its simplicity as a first approach for assessing whether this task could be solved automatically. The results should be only interpreted as a preliminary analysis suggesting that the proposed features do indeed contain useful information about the task. In contrast, a set of 10 expert human listeners reached very low agreement when solving this same task, highlighting the inherent difficulty of the problem. 

The results mentioned above were obtained on a subset of speakers selected for having few errors by the VA. When evaluating on the rest of the speakers, we found that the system did not perform better than random. We believe this happens due to a combination of less controlled environments, in the case of the sessions recorded through the internet, smaller amounts of data per sample before the first error by the VA, and subjects that were less committed to the task, reporting smaller differences in scores across conditions. 
Further data collection is needed to confirm the findings in this paper in a less controlled setting with larger amounts of subjects and more speech per subject. We plan to start the work of collecting a new dataset in the near future, using a novel protocol designed to engage the subjects using a game scenario.

Finally, several interesting questions arise from these experiments. Why are some speakers more affected by the initial score than others? Could this behavior be predicted from their personality traits or from their prior familiarity with VAs? Are some phrases more prone to contain useful information about the condition than others? We will address these and other questions in future research. 

\section{Acknowledgements}
This material is based upon work supported by the Air Force Office of Scientific Research under award number FA9550-18-1-0026. We also gratefully acknowledge the support of NVIDIA Corporation for the donation of a Titan Xp GPU.

\bibliographystyle{elsarticle-harv} 

\bibliography{mybib.bib}

\end{document}